\documentclass[iop]{emulateapj}
\usepackage{float}
\newcommand{\rb}[1]{\raisebox{1.5ex}[-1.5ex]{#1}}

\shorttitle{Discovery of a Magellanic galaxy at $z$=0.12}
\shortauthors{A. Koch et al.}

\begin{document}

\title{Major mergers with small galaxies -- \\the discovery of a Magellanic-type galaxy at \lowercase{$z$}=0.12\altaffilmark{$\dagger$}}

\author{
Andreas Koch\altaffilmark{1,2},  
Matthias J. Frank\altaffilmark{1},
Anna Pasquali\altaffilmark{3},
R. Michael Rich\altaffilmark{4,2}, 
\and
Andreas Rabitz\altaffilmark{5}
}
\altaffiltext{1}{Zentrum f\"ur Astronomie der Universit\"at Heidelberg,  Landessternwarte, K\"onigstuhl 12, 69117 Heidelberg, Germany}
\altaffiltext{2}{Visiting Astronomer, Kitt Peak National Observatory, National Optical Astronomy Observatory, which is operated by the Association of Universities for 
Research in Astronomy (AURA) under cooperative agreement with the National Science Foundation.}
\altaffiltext{3}{Zentrum f\"ur Astronomie der Universit\"at Heidelberg,  Astronomisches Rechen-Institut, M\"onchhofstrasse 12, 69117 Heidelberg, Germany}
\altaffiltext{4}{University of California Los Angeles, Department of Physics \& Astronomy, Los Angeles, CA, USA}
\altaffiltext{5}{Leibniz-Institut f\"ur Astrophysik Potsdam, An der Sternwarte 16, D-14482 Potsdam, Germany}
\email{akoch@lsw.uni-heidelberg.de}
\altaffiltext{$\dagger$}{Based on data obtained at the WIYN facility and the Large Binocular Telescope (LBT). 
The LBT is an international collaboration among institutions in Germany, the United States, and Italy. 
LBT Corporation partners are: 
LBT Beteiligungsgesellschaft, Germany, representing Heidelberg University, 
the Max-Planck Society, and the Leibniz-Institut f\"ur Astrophysik Potsdam. The University of Arizona on behalf of the Arizona university system; Istituto Nazionale di Astrofisica, Italy; The Ohio State University, and The Research Corporation, on behalf of The University of Notre Dame, University of Minnesota and University of Virginia.}
\begin{abstract}
We report on the serendipitous discovery of a star-forming galaxy at redshift $z$=0.116 with morphological features that indicate an ongoing merger.
This object exhibits two clearly separated components with significantly different colors, plus a possible tidal stream. 
Follow-up spectroscopy of the bluer component revealed
a low star-forming activity of 0.09 M$_{\odot}$\,yr$^{-1}$ and a high metallicity of 12+log(O/H)=8.6. 
Based on comparison with mass-star-formation-rate and mass-metallicity relations, 
and on fitting of spectral energy distributions, we 
obtain a stellar mass of  3$\times 10^9$ M$_{\odot}$, 
which renders this object  comparable to the Large Magellanic Cloud (LMC).  
{ Thus our finding provides a further piece of evidence of a major merger already acting on small, dwarf galaxy-like scales.} 
\end{abstract}
\section{Introduction}
Minor and major mergers are an important source for the hierarchical build-up of large-scale structures such as 
galaxies. These processes have been prominently traced over a large portion of the Universe, 
from our own Milky Way (MW; ingesting the Sagittarius dwarf galaxy; Ibata et al. 1994) to the immediate vicinity of M 31 (in the form of the Giant Stellar Stream; Ibata et al. 2001; McConnachie et al. 2009), 
and also in the Local Universe (Mart\'{\i}nez-Delgado et al. 2010; Rich et al. 2012). 
In all those cases, all participants in the mergers are still resolvable into stars, allowing for detailed studies of the underlying stellar populations. 
Similarly, dramatic galaxy interactions are observed  at larger distances (e.g., Forbes et al. 2003; Conn et al. 2011) 
and on the scales of galaxy clusters  (Koch et al. 2012). Above redshifts of $z\ga 1$, the major merger rate appears to diminish (e.g., Williams et al. 2011).

It remains an intriguing question, down to which scales mergers will play an important role. 
For instance, there are indications that Fornax, the second-most massive dwarf spheroidal satellite to the MW, has experienced several major 
accretion events of gas-rich systems  (Hendricks et al. 2014). Moreover, the Large and Small Magellanic Clouds (LMC and SMC, respectively) 
are likely to have had repeated, close interactions, triggering star formation (Harris \& Zaritsky 2009)
and shaping the warped bar and irregular structure of the LMC and the associated, gaseous Magellanic Stream (Besla et al. 2012).
Based on their orbital characteristics, it has been suggested that both irregular dwarfs had entered the MW system as a binary pair (Kallivayalil et al. 2012). 

However, galaxy pairs like the LMC-SMC-MW system are not overly common in the Local Universe, nor are they predicted in simulations of hierarchical accretion: 
The occurrence of satellites with a mass of the LMC or above to MW-sized galaxies lies at a mere 12\% (Liu et al. 2011; Tollerud et al. 2011; Robotham et al. 2012), 
consistent with the Millennium simulations of  Boylan-Kolchin et al. (2011). Furthermore, the LMC is unusual in that it has a remarkably blue color compared to 
its peers, both in galaxy groups  and in isolation, possibly owing to the merger-induced star formation (Tollerud et al. 2011; Besla et al. 2012). 

Given the importance of baryon removal from galaxies during such interactions  and the ensuing morphological transformations in the form of 
extended, gaseous bridges and tails,   
Magellanic-type irregular galaxies are good candidates to be targeted in H\,{\sc I} surveys. 
Likewise, many spectroscopic and imaging surveys exist to investigate the star-forming nature of merging and regular galaxies,  
and to study the evolution of galaxies from the Local Universe out to large redshifts, 
connecting  stellar masses with the star-formation and gas and dust properties as well as with  morphology (e.g., Overzier et al. 2009; Driver et al. 2009; Eales et al. 2010; van der Wel et al. 2011). 

Here we report on the serendipitous discovery of a low-mass galaxy which shows clear evidence of an ongoing merger event. 
This paper is organized as follows: in \S2 we introduce the discovery imaging data and spectroscopic follow-up. \S3 describes the photometric characterization, while \S4 is 
dedicated to the spectroscopic measurements in terms of star formation rate (SFR), metallicity, and stellar mass estimates. These are 
then discussed in \S5, where we draw a parallel to the properties of the LMC, before summarizing our findings in \S6. 
\section{Data}
\subsection{WIYN images}
We detected the galaxy, J021904.69+200615.4 (shortened to J021904 in the following),  
first on imaging of the Segue~2 ultrafaint dwarf spheroidal (cf. Belokurov et al. 2009)
we obtained with the 3.5-m Wisconsin-Indiana-Yale-NOAO (WIYN\footnote{The WIYN Observatory is a joint facility of the University of Wisconsin-Madison, Indiana University, Yale University, and the National Optical Astronomy Observatory.}) 
telescope on Kitt Peak
on Oct. 07, 2010.
Here, we used the MIMO mosaic camera, which provides a large field of view of 
$9.6\arcmin\times9.6\arcmin$ spread over two detectors, each with a pixel scale of 0.14$\arcsec$. 
Our exposure times were 6$\times$300 s in each V and I and 9$\times$300 sec in B, with individual exposures dithered across the gaps.
The seeing was 1$\arcsec$ on average.

Our standard data reduction was carried out in IRAF. 
The field around Segue 2 is covered by the Sloan Digital Sky Survey (SDSS; Alam et al. 2015). 
Thus, to photometrically calibrate our data, we used SExtractor (Bertin \& Arnouts 1996) to obtain aperture photometry 
and aperture corrections
 for stars in the stacked B,V, and I images. 
The instrumental magnitudes of $\sim$240 stars were then matched against the SDSS DR12 
photometric catalog that had been transformed to the Johnson-Cousins BVI system using the photometric transformations of Jordi et al. (2006).
For reference, Table~1 lists the photometry of the extended galaxy J021904  { listed in the SDSS.
In the SDSS, the total magnitudes were obtained from the best-fit exponential profile, while a de Vauculeurs profile yielded 
low likelihoods. These magnitudes in the SDSS are based on integrating this profile out to three scale-radii, or $\sim12\arcsec$ (see also  Sect.~3.1), 
which full captures the flux from both merger components.}
\begin{center}
\begin{deluxetable}{cc}[b!]
\tabletypesize{\scriptsize}
\tablecaption{Photometric characteristics of J021904 from the SDSS.}
\tablewidth{0pt}
\tablehead{ \colhead{ Filter}  & \colhead{ Magnitude\tablenotemark{a}} }
\startdata
u & \phantom{$-$}20.84$\pm$0.24\\
g & \phantom{$-$}19.63$\pm$0.04  \\
r & \phantom{$-$}18.89$\pm$0.04  \\
i & \phantom{$-$}18.37$\pm$0.03  \\
z & \phantom{$-$}17.99$\pm$0.08   \\
M$_g$ & $-20.53\pm$0.15\tablenotemark{b}
\enddata
\tablenotetext{a}{ Total magnitude from SDS, obtained from an exponential profile fit.}
\tablenotetext{b}{Adopting the luminosity distance of 559 Mpc.}
\end{deluxetable}
\end{center}

In Fig.~1 we show the intensity maps from our BVI imaging. 
J021904 was discovered as an amorphous, extended system showing clear signs of a double core structure, 
most prominently on the B-image, and a possible tidal tail emerging on the SouthEast side
of the galaxy, indicative of an ongoing merger event (Fig.~1). 
Comparison with the MMT images of Belokurov et al. (2009) and the SDSS images indicates that the features in question, while less pronounced on those shallower images, 
 are real and not artefacts of our imaging or reduction procedures. 
\begin{figure}[htb!]
\begin{center}
\includegraphics[angle=0,width=0.8\hsize]{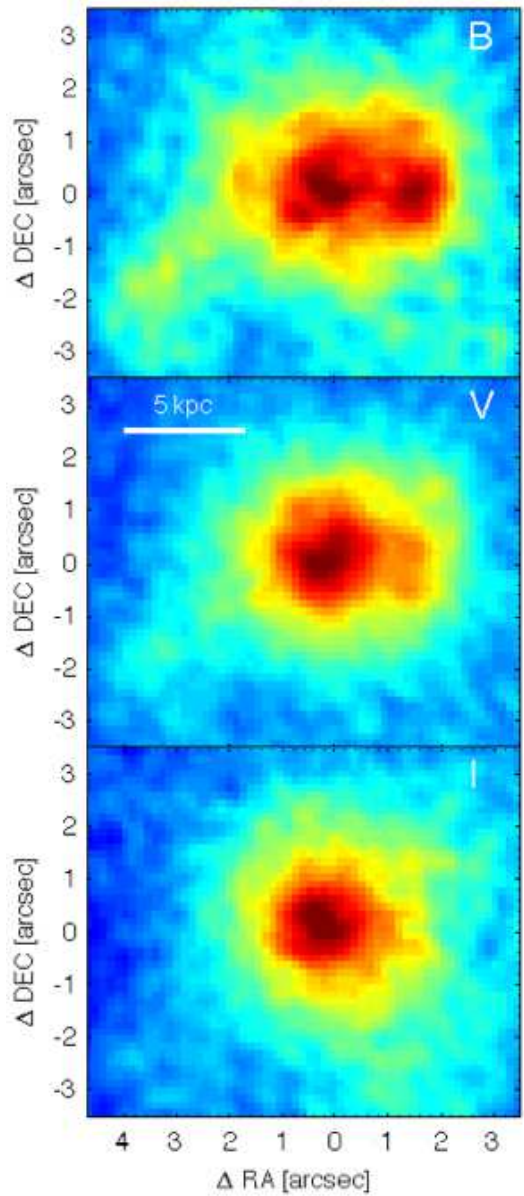}
\end{center}
\caption{Top to bottom: B, V, and I-band intensity maps of the galaxy J021904. 
Two components are clearly visible in the B-band image. The images have been smoothed with a Gaussian kernel of 1$\arcsec$. A scale bar of 5 kpc is indicated. North is up, East is left.}
\end{figure}

It is noteworthy that the Eastern component is much redder than the Western one, which we highlight in Fig.~2 in terms of the B$-$I color map of J021904. 
The entire object has a  visual, radial extent of $\sim 8\arcsec$ (Figs.~1,2; see also Sect.~3), where we 
note that the SDSS lists a Petrosian radius for J021904 of 7.1$\pm$3.8$\arcsec$. 
\begin{figure}[htb]
\begin{center}
\includegraphics[angle=0,width=0.9\hsize]{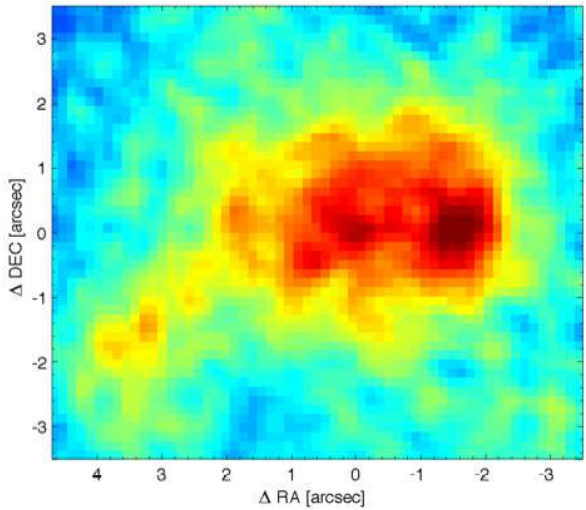}
\end{center}
\caption{B$-$I color map { of the same region as in Fig.~1}, highlighting the color difference between the two components.
{ This map have been smoothed with a Gaussian kernel of 1$\arcsec$.}}
\end{figure}
\subsection{MODS spectroscopy}
We obtained follow-up spectroscopy with the first Multi-object Double Spectrograph (MODS1; Pogge et al. 2010) at the Large Binocular Telescope (LBT) on Nov. 18, 2012. 
While four slit masks for multi-object spectroscopy had been designed so as to representatively cover many subcomponents of the target galaxy, only one 
could be observed due to unfortunate weather conditions. 
Here, the slit was placed on the Western, bluer component of the galaxy (Fig.~3). 
\begin{figure}[htb]
\begin{center}
\includegraphics[angle=0,width=0.95\hsize]{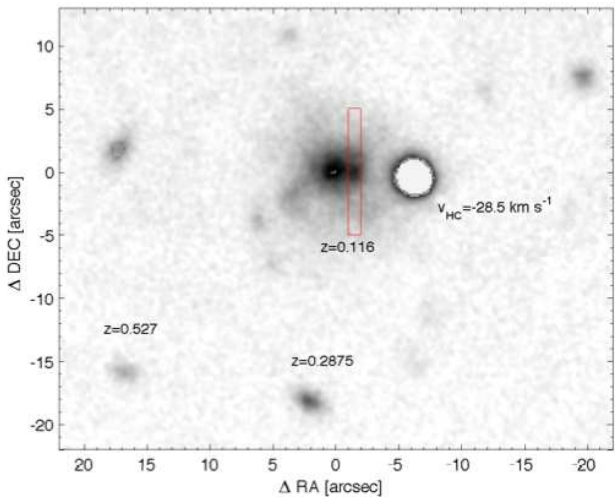}
\end{center}
\caption{Combined BVI intensity map of the surroundings of J021904. A red rectangle indicates the placement of the MODS slit. Three additional objects with derived redshift 
{ $z$ and stellar heliocentric radial velocity, v$_{\rm HC}$, }
from our spectroscopy are also labeled.}
\end{figure}

In practice, we used a single $1\arcsec\times10\arcsec$ slit and  exposed  for 5$\times$1800 s to facilitate cosmic ray removal. 
The chosen setup resulted in a full spectral coverage of 3000--9000\AA~ 
at a  resolution of $R=1300$ as 
measured from the sky lines. 
We used the \textsc{modsCCDRed} package\footnote{\textsc{modsCCDRed} by R.~W.~Pogge, 
available at \\ \url{http://www.astronomy.ohio-state.edu/MODS/Software/modsCCDRed/}} for basic
 data reduction, i.e. bias-removal, and applying pixel-flats.
Since small shifts in spatial and wavelength directions between individual exposures were present, 
we opted to reduce the individual science frames independently.
For this, cosmic rays were identified using a python variant\footnote{\textsc{cosmics.py} by M.~Tewes, 
available at \\ \url{http://obswww.unige.ch/$\sim$tewes/cosmics\_dot\_py/}} 
of the L.A. Cosmic algorithm (van Dokkum 2001). We then used the \textsc{modsIDL} 
pipeline\footnote{\textsc{modsIDL} by K.~V.~Croxall \& R.~W.~Pogge, available at \\  \url{http://www.astronomy.ohio-state.edu/MODS/Software/modsIDL/}} to  
trace slits, derive the wavelength calibration from day-time arc lamp exposures, perform sky-subtraction on the 
two-dimensional images (using the method of Kelson 2003), and to extract one-dimensional spectra. 
Absolute flux calibration was reached via exposure of the spectrophotometric standard star GD\,71.
For each slit, the spectra extracted from the individual exposures were then combined using an 
inverse-variance weighted mean. As a result, the signal-to-noise ratio in the continuum near H$\alpha$ is $\sim$10 per pixel.

The placement of additional slits in the immediate surroundings of the galaxy of interest also allowed us to characterize 
several fore- and background objects (labeled in Fig.~3). Two of these were identified as background galaxies at 
z=0.527 (with prominent Ca\,{\sc ii} HK absorption) and z=0.288 (an emission line galaxy). 
Likewise, the  object West of J021904 (at $g$=19.4 mag) has a { heliocentric radial velocity
of v$_{\rm HC}\,=\,-28$ km\,s$^{-1}$. 
This compares to the systemic velocity of the Segue 2 dwarf galaxy of $-40$ km\,s$^{-1}$ and its internal  velocity dispersion of 3.4 km\,s$^{-1}$ (Belokurov et al. 2009),  
which  renders this object a clear foreground star.}  
%
%
%
%
\section{Photometric measurements}
As our surface brightness measurements on the center of either component of the galaxy indicate, 
both sides have comparable brightnesses in the B-band, while the Western part is considerably fainter, by $\sim$0.6 mag/$\sq\arcsec$, 
in the I-band (see Table~2). 
\begin{center}
\begin{deluxetable}{ccccc}[H]
\tabletypesize{\scriptsize}
\tablecaption{Derived photometric parameters of J021904.}
\tablewidth{0pt}
\tablehead{ \colhead{Parameter}  & \multicolumn{3}{c}{Value} & \colhead{Notes} \\
\cline{2-4}
\colhead{}  & \colhead{B} & \colhead{V} & \colhead{I} & \colhead{} }
\startdata
$\mu_{\rm E}$ [mag/$\sq\arcsec$]  & 24.34$\pm$0.08  & 23.44$\pm$0.12 &  22.94$\pm$0.34& 1 \\
$\mu_{\rm W}$  [mag/$\sq\arcsec$] & 24.52$\pm$0.08  &  23.73$\pm$0.13  &  23.47$\pm$0.36  & 2 \\
r$_e$ [$\arcsec$] & 4.1$\pm$1.7  & 3.7$\pm$0.9 &  3.5$\pm$1.2 & 3 \\
 C & 3.6 & 3.6 & 3.5 & 4 \\
A & 0.15 & 0.17 & 0.13 & 4 \\
S & 0.13 &  0.28 & 0.16 & 4
\enddata
\tablecomments{(1) Central surface brightness of the Eastern (redder) component.
(2) Central surface brightness of the Western (bluer) component. (3) { Effective radius of the exponential profile fitted to the 
entire galaxy.} { (4) Concentration, Asymmetry, and Smoothness, as defined by 
Conselice et al. (2000).}}
\end{deluxetable}
\end{center}
Using Herschel observations of local  dwarf galaxies, 
Clark et al. (2015) found a large fraction of objects with remarkably blue colors (albeit in FUV$-K_S$). 
Similar to J021904, those  galaxies show an  irregular  morphology. Furthermore, Clark et al. (2015) pointed out that those, overall bluer galaxies are 
 dominated by dust and that they are the  most actively star forming galaxies of their survey. 
This could indicate that the Western, bluer component (albeit in the optical bands considered here) of the merger we observed was contributed by a galaxy with a larger dust fraction than 
the other partner. 

The separation between the centers of the two blobs is 1.6$\arcsec$, corresponding to 3.5 kpc at the distance of J021904 (Sect.~4.1).
 Several knots are detected, at the 1$\sigma$-level above the background, towards the SouthEast on the B$-$I map at projected distances of 7.5, 9, and 10.3 kpc, 
 but at the low spatial resolution of the images it remains inconclusive if they are localized star forming regions, 
tidal streams, or   accreting dwarf galaxies. 
Stellar and/or gaseous streams in the Local Group reach lengths of $>$140 kpc (e.g., Ibata et al. 2001), which, depending on the inclination angle to the observer, 
could be seen as similar features to what we observe in our target galaxy. 
Likewise, dwarf satellites to Magellanic galaxies at projected distances of $\sim$10 kpc are common (e.g., Rich et al. 2012) 
so that the observed knots could in fact be the relics of an accreted dwarf galaxy. 
\subsection{Radial profiles}
In Fig.~4 we show the azimuthally averaged surface brightness profiles of J021904 in each of the targeted filters. To this end, we have carefully masked out the near-by star (cf. Fig.~3).  
The best-fit exponential profiles are also shown, yielding consistent effective radii within the errorbars (Table~2). 
{ For consistency with the SDSS (Sect.~2.1), we use the parameterization 
I(R)\,=\,$\exp\left(-1.68\,{\rm r}/r_e\right)$
in the following, where  $\mu=-2.5\log$\,I 
and $r_e$ denotes the effective scale radius containing half of the light. 
}
While the difference is not statistically significant, we nonetheless point out that 
the radius obtained in the B-band is in fact the largest, owing to the contribution from the secondary merger component in the brightness distribution. The radii thus obtained correspond to 
{ 7.6--9.1} kpc at the distance of the galaxy (Sect.~4.1).  We also attempted Sersi\'c fits, but the resulting indices of 0.98--1.05 are fully consistent with an exponential light distribution in this galaxy, 
also in line with the preponderance of disk-like main sequence galaxies at higher redshifts (Wuyts et al. 2011). 
{ Furthermore, avoiding the seeing disk by restricting the fits to $r>1\arcsec$ had only a negligible influence on the results in that the best-fit radii increased by less than 5\% when using the outer regions. 
only. This is much smaller than the uncertainties on the best-fit parameters (Table~2).}
\begin{figure}[tb]
\begin{center}
\includegraphics[angle=0,width=1\hsize]{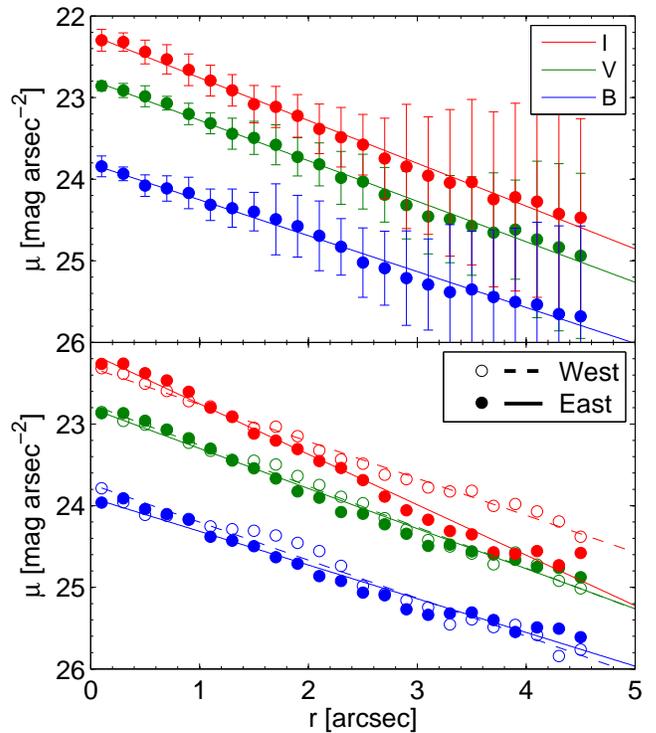}
\end{center}
\caption{Azimuthally averaged surface brightness profiles of J021904 from our B,V, and I images (top panel). 
The bottom panel shows the profiles when restricting to the Western ($\xi\le0$) and Eastern ($\xi>0$) halves. 
Best-fit exponential profiles are indicated for each subset.
}
\end{figure}

In order to look for systematic differences between either component of the double-peaked galaxy, we divided our images into an Eastern and a Western component and show the resulting
profiles in the bottom panel of Fig.~4. 
{ Also here we tested the effects of seeing by using only data beyond 1$\arcsec$ and found a 8\% (2\%) increase (decrease) in the scale radii in the Eastern (Western) halves.} 
Here it is noteworthy that the profiles in the V-band are essentially indistinguishable. While the existence of the second brightness peak is clearly visible between 1--2$\arcsec$ in the 
blue, Western profile, the formal best-fit radius is larger on the Eastern side by 0.3$\arcsec$, owing to the presence of the purported tidal extension. 
In contrast the faintness of these features in the reddest filters renders the scale radius of J021904 considerably smaller, by { 1.1$\arcsec$}. 
Thus the accreting system or tidal tail could indicate a merger with a bluer object or ongoing star-formation in localized regions, possibly triggered by the merger. 
\subsection{ CAS parameters}
{ 
In order to further parameterize the morphology and extent of the disturbances of J021904 we determined the ``CAS'' parameters (Conselice 2003), viz., 
the galaxy's concentration (C), asymmetry (A), and Clumpiness (S), which are also known to correlate with galaxy type and/or color. 
Here, we followed the formalism of Conselice et al. (2000) in measuring these parameters from our images. 

\vspace{1ex}
{\em (1) Concentration:}
C is defined as the logarithmic ratio of those radii that contain 80\% and 20\% of the cumulative flux contained within a certain Petrosian radius, more specifically, within
$1.5\times r(\eta=0.2)$, where $\eta$ denotes the ratio of the local surface brightness to the average surface brightness within radius $r$ (e.g., Djorgovski \& Spinrad 1981; Kent 1985; Bershady et al. 2000). 
Thus, 
\begin{equation}
{\rm C}\,=\,5\times\log\,(r_{ 80\%}/r_{ 20\%}). 
\end{equation}
For our galaxy we found C$=(3.6,3.6,3.5)$ in B,V, and I, indicating similar concentrations of the galaxy irrespective of the filter.  
Typical values of C range from 2--5, where C$>$4 is mainly found for elliptical, while disk galaxies have concentrations between 3 and 4. 

\vspace{1ex}
{\em (2) Asymmetry:}
Galactic asymmetries can be obtained by rotating the image, centered on the galaxy,  by 180${\degr}$ and subtracting 
this rotated image from the original one. In practice, this is quantified by 
\begin{equation}
{\rm A}\,=\,\min\left( \frac{\Sigma\,| I_{\rm rot}\,-\,I_0 |}{\Sigma\,| I_0 |}\right)\,-\,\min\left(\frac{\Sigma\,| B_{\rm rot}\,-\,B_0 |}{\Sigma\, | I_0|} \right). 
\end{equation}
Here, $I_{\rm rot}$ and I$_0$ refer to the rotated and original images, respectively, and 
the summation over the fluxes is carried out within $1.5\times r(\eta=0.2)$ as above. $B$ denotes
the analogous contributions from a background field at sufficiently large distances that is free from real galactic structures 
and is introduced to account for contributions from uncorrelated noise. Finally, the minimization was employed so as to 
optimize the galaxy centroid  (Conselice et al. 2000). 
The asymmetry of J021904 was thus found to be A$=(0.15,0.17,0.13)$ in B, V, and I. 
Given the above definition of A, values around zero would indicate symmetry, whereas an A of unity is caused by 
large, asymmetric structures. It is noteworthy that there is no significant difference in A between the three different filters we investigated, which 
could be expected if localized star formation in the galaxy would incur varying asymmetry in different bands. 
 
\vspace{1ex}
{\em (3) Clumpiness:} 
High-frequency variations in a galaxy, i.e., any form of clumpy substructures, can be 
investigated by subtracting a smoothed version of the image from the original one. Following Conselice (2003), we smoothed our images, $I_0$ with a kernel of width 0.3$\times r(\eta=0.2)$ (resulting in $I_{\sigma}$
and accounted for the background noise by introducing the same term $B_0$ as above (eq.~2).
\begin{equation}
{\rm S}\,=\,10\times\left[\left( \frac{\Sigma\,I_{0}\,-\,I_{\sigma} }{\Sigma\, I_0}\right)\,-\,\left(\frac{\Sigma\,B_{0}\,-\,B_{\sigma}}{\Sigma\,  I_0} \right)\right]. 
\end{equation}
Here, we found S=(0.13,0.28,0.16) from the B, V, and I images of J021904.

In Fig.~5 we put our measurements into context by using the galaxy samples of Conselice (2003), who established a new classification scheme of galaxies based on these CAS parameters. 
While the literature values are based on R-band images, we nonetheless overplot our results from the V-images, considering that the variations of CAS between our three filters was found negligible. 
\begin{figure}[tb]
\begin{center}
\includegraphics[angle=0,width=1\hsize]{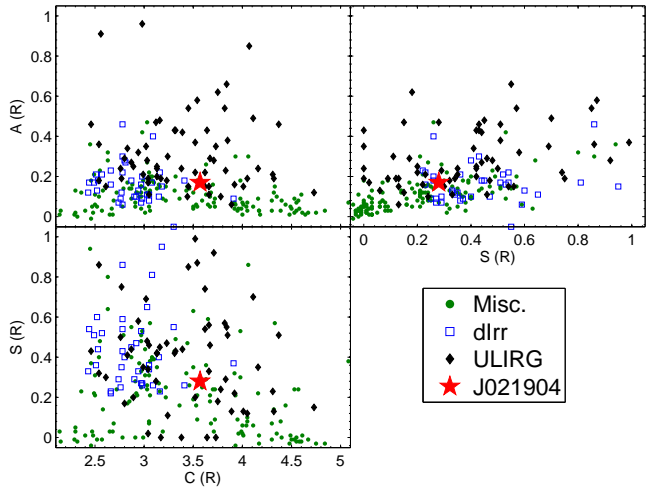}
\end{center}
\caption{ Concentration, Asymmetry, and Clumpiness parameters of various galaxy types from Conselice (2003). Our measurement for J021904 from the V-image is indicated as a red star symbol. 
Under ``Misc.'' we grouped Conselice's (2003) samples containing giant ellipticals/S0s, early- and late-type spirals, and starburst- and dwarf elliptical galaxies.}
\end{figure}
Despite its visually irregular nature, the asymmetry of J021904 is moderate and only a few of the dIrrs from Conselice (2003) fall into the same parameter space as our galaxy. 
A  strong overlap is found for regular galaxies, such as edge-on disks, which, however, can be refuted by the visual appearance on our images. 
There are, however, also several ULIRGs found at similar parameter combinations; these show, on average, the largest asymmetries due to the active evolution 
of these systems, often correlated with ongoing major merger activities. While not extreme in CAS,  J021904's asymmetry and clumpiness is consistent with such a merger notion. 

Finally, it is also important that Conselice et al. (2000) note that objects will appear less asymmetric when poorly resolved. 
At the large distance and small apparent size of the galaxy in question,  it is possible that the asymmetry is underestimated from our images, placing it further into the regime 
of interacting galaxies. 
}
%
%
%
%
\section{Spectroscopic measurements}
Fig.~6 shows the flux calibrated spectrum, shifted to the rest-frame.
Several strong emission lines are clearly seen and we identified ten lines that were significantly measurable above the noise. 
\begin{figure}[htb]
\begin{center}
\includegraphics[angle=0,width=1\hsize]{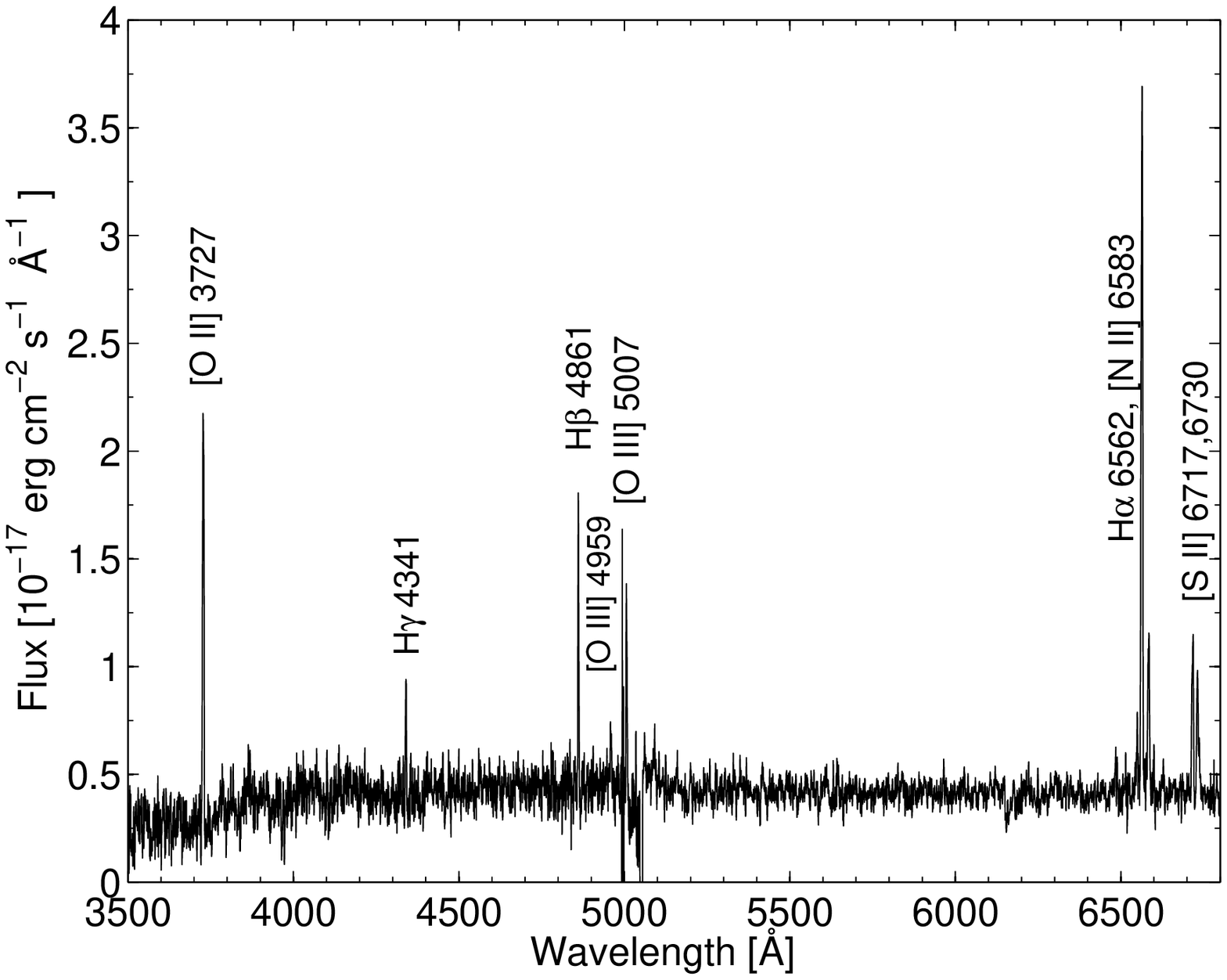}
\includegraphics[angle=0,width=1\hsize]{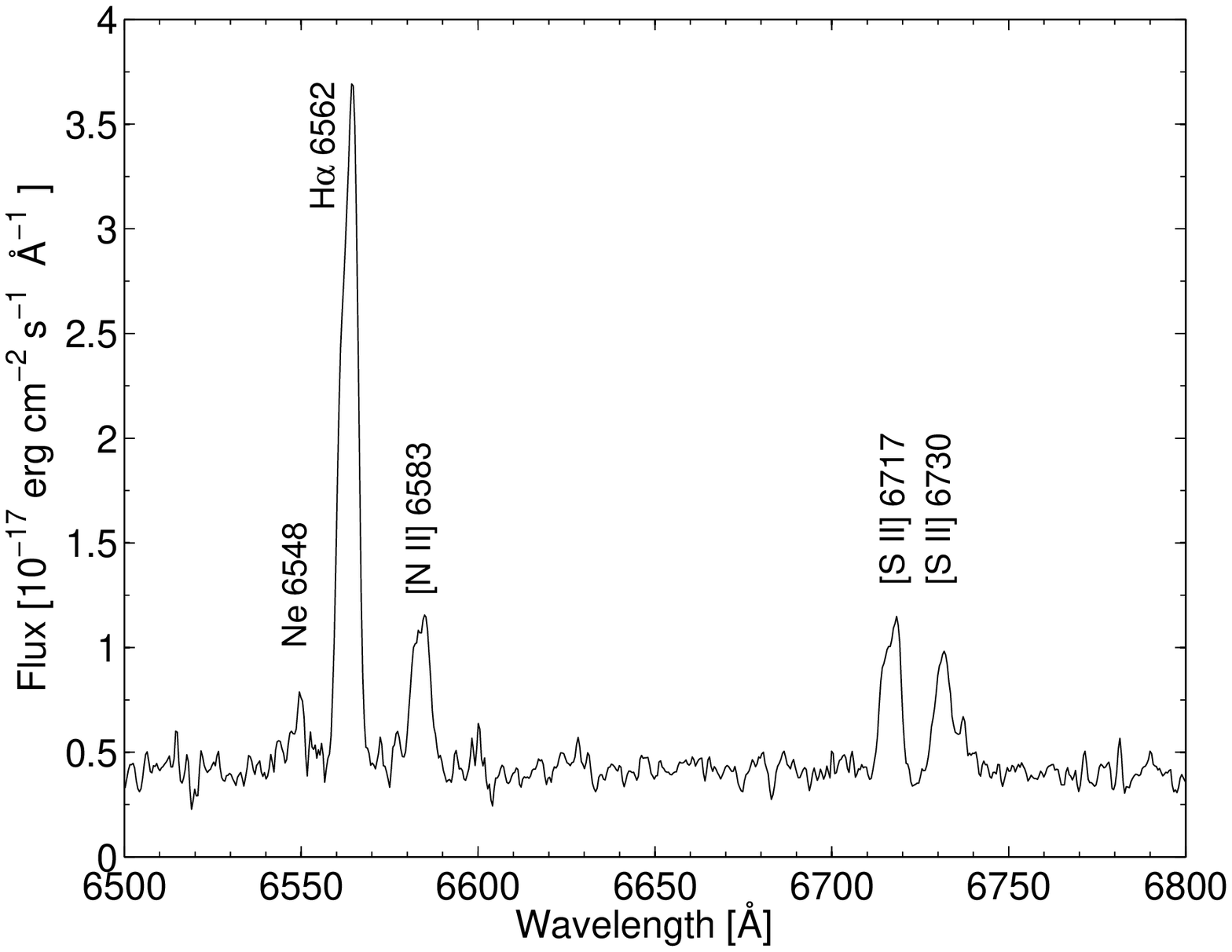}
\end{center}
\caption{Flux-calibrated  MODS spectrum of J021904, converted to the rest-frame. The bottom panel shows a zoom around the H$\alpha$, [N II]\,$\lambda$6584, and the [S II]\,$\lambda\lambda$6717,6730 features.
Note the asymmetries in these strong lines. }
\end{figure}
The first thing to note is
that all lines show an asymmetry or are even split into two components as in the case of the [O III]\,4959 line. 
Using IRAF's {\em splot} task we de-blended the lines and could measure a systematic split of 
142$\pm$6 km\,s$^{-1}$, which we will further discuss in Sect.~5. 
While the resolution of our spectra is sufficient to detect the asymmetries in the line shifts, it is not enough to accurately deconvolve the lines into two separate components 
and to reliably determine  individual line fluxes. 
Thus we continue with measuring the flux via numerical integration of the emission lines from the  spectra after correction for local continuum.
The errors on the observed fluxes, in turn,  were estimated from the variance spectrum via repeated integration of the lines in a Monte Carlo sense. 
\subsection{Redshift and distance}
Based on the shifts of  ten prominent emission lines, determined from Gaussian fits using  the {\em splot} task,  
 we obtained a redshift for J021904  of 0.116472$\pm$0.000049. 
This value is in excellent agreement with the photometric redshift from the SDSS of 0.132$\pm$0.031.
Adopting a Hubble constant of H$_0$=67.8$\pm$0.9 km\,s$^{-1}$\,Mpc$^{-1}$ and the cosmology derived by the Planck collaboration (2015) this corresponds to 
a luminosity distance of 559$\pm$7 Mpc (Wright 2006); in turn, 1$\arcsec$ equals to 2.2 kpc at this distance
and the pixel size on our WIYN images is  305 pc. 

\subsection{Extinction, electron temperature, and density}
In order to derive the entire color excess due to dust extinction, $<$E$_g$(B$-$V)$>$, 
averaged over the slit, we followed 
 Pasquali et al. (2011) in measuring the Balmer decrement in terms of the observed flux ratio of the  
H$\alpha$ and H$\beta$ lines. 
Since there is only marginal evidence for stellar absorption around the H$\gamma$ and H$\delta$ lines,  
the underlying stellar absorption was deemed negligible. 
Thus we determine 
\begin{equation}
<{\rm E_g(B-V)}>\,=\,-\frac{\log\,(\,R_{\rm \,obs}\,/\,R_{\rm \,int}\,)}{0.4\,\left[\,\kappa\,(\lambda_{\rm H\alpha})-\kappa\,(\lambda_{\rm H\beta})\,\right]}, 
\end{equation}
where $R_{\rm obs}$ and $R_{\rm int}$ refer to the observed and intrinsic line flux ratios, H$\alpha$/H$\beta$. 
For the theoretical value we assume $R_{\rm int}$=2.86 for case B recombination (Osterbrock \& Ferland 2006).
Moreover,  $\kappa(\lambda)$ was evaluated  from the extinction
law of Fitzpatrick (1999) at the respective wavelengths of the Balmer lines (see also Calzetti 2001). 
As a result, we found a total, i.e., intrinsic plus foreground, reddening of 0.43$\pm$0.04 mag. 
Since the Schlafly \& Finkbeiner (2011) maps indicate a Galactic foreground reddening of E$_{\rm G}$(B$-$V)=0.19 mag, this leaves
an intrinsic $<$E$_g^i$(B$-$V)$>$ of 0.24 mag.
The overall large extinction (note that A$_V$=1.33 mag) is likely even underestimating the true value, since such large amounts of dust 
will attenuate further flux from the inner parts of the galaxy, disabling their use for our extinction estimates. 

Finally, the observed emission line fluxes were de-reddened via 
\begin{equation}
F_0\,(\lambda)\,=\,F_{\rm obs}\,(\lambda)\,\times\,10^{\,0.4\,<{\rm E_g(B-V)>}\,\kappa\,(\lambda)} 
\end{equation}
and we list these values in Table~3. 
\begin{center}
\begin{deluxetable}{cc}
\tabletypesize{\scriptsize}
\tablecaption{Emission line fluxes}
\tablewidth{0pt}
\tablehead{ \colhead{}  & \colhead{Corrected flux $F_0$} \\
\colhead{\rb{Feature}} & \colhead{[10$^{-17}$\,erg\,cm$^{-2}$\,s$^{-1}$]}}
\startdata
$[$S II$]$\,$\lambda$6730 & 9.27$\pm$0.17\\
$[$S II$]$\,$\lambda$6717 & 9.76$\pm$0.17\\
$[$N II$]$\,$\lambda$6584 & 12.20$\pm$0.18\\
H$\alpha$\,$\lambda$6563 & 46.53$\pm$0.15 \\
Ne\,$\lambda$6548 & 3.64$\pm$0.10\\
$[$O III$]$\,$\lambda$5007 &  11.28$\pm$1.02\\
$[$O III$]$\,$\lambda$4959 & 5.01$\pm$0.87\\
H$\beta$\,$\lambda$4861 & 16.27$\pm$0.70 \\
$[$O III$]$\,$\lambda$4363 & $<$0.51\tablenotemark{a}\\
H$\gamma$\,$\lambda$4340 & 6.29$\pm$0.96 \\
H$\delta$\,$\lambda$4102& 2.7:\tablenotemark{b} \\
$[$O II$]$\,$\lambda$3727 & 58.52$\pm$1.51
\enddata
\tablenotetext{a}{1$\sigma$ upper limit.}
\tablenotetext{b}{Uncertain value due to continuum placement.}
\end{deluxetable}
\end{center}

Since the auroral [O III] line at 4363\AA~is too weak to be seen in the MODS spectrum, 
we cannot obtain  a direct estimate of the electron temperature, $T_e$, of the gas phase 
from the common [O III] I($\lambda$4959+$\lambda$4959)/I($\lambda$4363) indicator.
Likewise, no other high-excitation, $T_e$-sensitive emission lines such as [O II] $\lambda$7320
or [N II] $\lambda$5755 could be significantly detected. 
We therefore conclude that the blue, Western component of J021904 
consists of a 
low excitation, metal-rich  region, where cooling can proceed via metal lines.  
Here, electron temperatures are typically found to be on the order of  $\sim$5000--7000 K.

For an estimate of the electron density, $n_e$, we employed the [S II]  I($\lambda$6717)/I($\lambda$6730) ratio
and the low $T_e$ adopted above, which yields a typical $n_e$ of 430$\pm$50 cm$^{-3}$.  
All spectroscopically derived parameters are listed in Table~3.
\begin{center}
\begin{deluxetable}{cc}
\tabletypesize{\scriptsize}
\tablecaption{Spectroscopically derived characterization of J021904}
\tablewidth{0pt}
\tablehead{ \colhead{Parameter}  & \colhead{Value} }
\startdata
$[$S II$]$ 6717/6731 & 1.05$\pm$0.03 \\
$[$O III$]$ (5007+4959)/4363 & $>$33.2\tablenotemark{a} \\
R$_{23}$ & 3.94$\pm$0.20 \\
O3N2 & 0.42$\pm$0.47 \\
$<$E$_g$(B$-$V)$>$ & 0.43$\pm$0.04 mag \\
$n_e$ & 430$\pm$50 cm$^{-3}$\\ 
$T_e$ & $<$23000 K\tablenotemark{a} / $\la$7000 K\tablenotemark{b}\\ 
12 + log(O/H) (from R$_{23}$) & 8.84$\pm$0.02 dex\\
12 + log(O/H) (from O3N2) & 8.59$\pm$0.15 dex\\
L(H$\alpha$) & (1.74$\pm$0.04)$\times$10$^{40}$ erg\,s$^{-1}$\\
SFR & (9.2$\pm$1.6)$\times$10$^{-2}$ M$_{\odot}$\,yr$^{-1}$\\
M$_{\ast}$ (from SFR) & (5.4$\pm$1.0)$\times$10$^8$ M$_{\odot}$ \\
M$_{\ast}$ (from O/H) & (3.9$^{+0.6}_{-0.5}$)$\times$10$^9$ M$_{\odot}$ \\
M$_{\ast}$ (from fundamental plane) & (2.1$^{+1.2}_{-0.8}$)$\times$10$^9$ M$_{\odot}$ \\
M$_{\ast}$ (from SED) & (4.3$^{+1.5}_{-1.0}$)$\times$10$^9$ M$_{\odot}$ 
\enddata
\tablenotetext{a}{1$\sigma$ limit based on the upper limit on the [O III]~4363 flux.}
\tablenotetext{b}{Low-excitation estimate based on the non-detection of the [O~III]~4363 line.}
\end{deluxetable}
\end{center}
\subsection{Oxygen abundance}
The presence of several strong emission lines in the spectrum of J021904 allowed us to estimate the metallicity of this galaxy from 
the O3N2 indicator (Alloin et al. 1979), defined as 
\begin{equation}
{\rm O3N2}\,=\,\log\,\frac{{\rm [O\,III]}\,\lambda 5007\,/\,{\rm H}\beta}{{\rm [N\,II]}\,\lambda6583\,/\,{\rm H}\alpha}. 
\end{equation}
Pettini \& Pagel (2004) devised an  empirical calibration of this ratio from a sample of  extragalactic HII regions
and yielded consistent results with a broad variety of other indicators, discussed, e.g., by 
Kewley \& Dopita (2002). Thus the oxygen abundance of J021904 is determined via 
\begin{equation}
12\,+\,\log({\rm O/H})\,=\,8.73\,-\,0.32\times{\rm O3N2}
\end{equation}
as 12+log(O/H) =  8.59$\pm$0.15, or [O/H] = $-0.07\pm0.15$ dex, adopting the  Solar oxygen abundance of 8.66 from Asplund et al. (2009). 
Since  this is an empirical calibration, our derived metallicity is rather insensitive to the adopted values for electron temperature and density (cf. Kewley \& Dopita 2002). 

Due to its wide application in the literature, we also employed the R$_{23}$ indicator, which reads
\begin{equation}
R_{23}\,=\,\frac{{\rm [O\,II]}\,\lambda 3727\,+\,{\rm [O\,II]}\,\lambda 4959}{{\rm H}\beta}. 
\end{equation}
Following the third-order polynomial calibration of this index to the metallicities of 53,000 galaxies by Tremonti et al. (2004), we 
obtain a higher oxygen abundance  for J021904 of 8.84$\pm0.02$ dex, bearing in mind discrepancies between the various oxygen abundance indicators (Kewley \& Ellison 2008)\footnote{Template 
fitting of the Ca\,{\sc ii} HK absorption lines 
with pPXF (Cappellari \& Emsellem 2004) yielded a consistent, overall metallicity, [M/H], of $-0.3$ dex. However, 
we did not pursue the determination of further population characteristics using this method, since all other, e.g.,  age-sensitive features such as the 
Mg-$b$ lines are not detectable in the spectrum.}. 
\subsection{Star formation rate}
In Fig.~7 we show the diagnostic diagram of Baldwin et al. (1981; hereafter BPT) for a sample of $\sim$180,000 star-forming  galaxies 
from the SDSS (Brinchmann et al. 2004). Here, using ratios of  emission lines close in wavelength efficiently reduces reddening effects. 
The BPT plot is then a powerful discriminator (e.g., Kewley et al. 2001), taking advantage of the fact that
stronger [N\,II] emission relative to the H$\alpha$ line is usually indicative of 
Active Galactic Nuclei (AGN) while stronger H$\alpha$ contribution suggests a starburst origin in star-formation galaxies.  
J021904 falls square on the main locus of star-forming galaxies in the BPT diagram, bolstering its earlier identification 
as a star-forming system. 
\begin{figure}[htb]
\begin{center}
\includegraphics[angle=0,width=1\hsize]{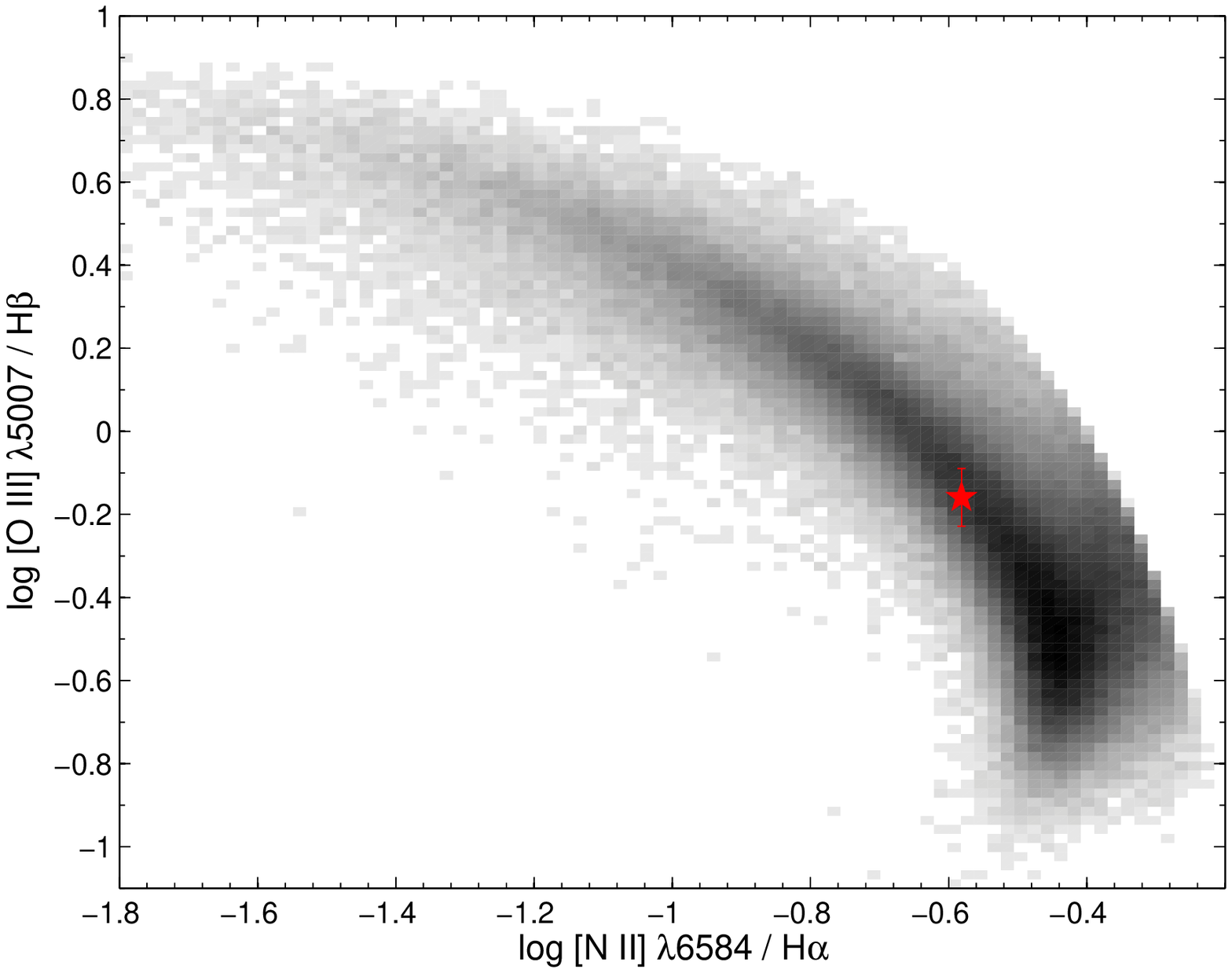}
\end{center}
\caption{BPT diagnostic diagram, using star-forming galaxies from the SDSS (Brinchmann et al. 2004). 
AGN have already been pruned from the sample. J021904  is indicated by a red star symbol.}
\end{figure}

From our spectrum we derive an H$\alpha$-luminosity of L(H$\alpha$)\,=\,(1.74$\pm$0.04)$\times10^{40}$ erg\,s$^{-1}$.
This can be converted into a star formation rate (SFR) using the empirical calibration of 
Calzetti et al. (2007), which reads 
${\rm SFR\,\,[M}_{\odot}$\,yr$^{-1}$]\,=\,5.3$\times$10$^{-42}$\,L(H$\alpha$)\,[erg\,s$^{-1}$] 
(see also Moustakas et al. 2006 for alternative parameterizations). 
Thus we find a SFR of (9$\pm$2)\,$\times$\,10$^{-2}$ M$_{\odot}$\,yr$^{-1}$ for the Western, blue component of J021904.
The quoted uncertainty accounts for the flux-measurement based error on the H$\alpha$-luminosity and a 20\% variation in the constant 
of the above relation, as suggested by Calzetti et al. (2007). 
This low value suggests that J021904 is forming stars at a moderate pace.
\subsection{Stellar mass}
There exists a wealth of (stellar-) mass-metallicity, mass-SFR, and metallicity-SFR relations for various types of galaxies (e.g., Lequeux et al. 1979; Tremonti et al. 2004), 
down to the faintest stellar systems known (e.g., Kirby et al. 2011; Koch \& Rich 2014).  
In the following we attempt several measures to estimate the stellar mass of J021904.

\vspace{1ex}
{\em (1) Stellar mass from SFR:} 
Fig.~8 shows the main sequence of star-forming galaxies from the SDSS (Brinchmann et al. 2004) in the SFR-stellar mass diagram, where the stellar masses
had been derived from the objects' photometry alone (e.g., Kauffmann et al. 2003). 
Note that this  sample is restricted to redshifts between 0.005$<z<$0.22 so that J021904  representatively lies  within this range.
Interpolating the fiducial line in this parameter space 
to the observed SFR for J021904, we obtain  a stellar mass of (5.4$\pm$1.0)$\times$10$^8$ M$_{\odot}$. 
\begin{figure}[htb]
\begin{center}
\includegraphics[angle=0,width=1\hsize]{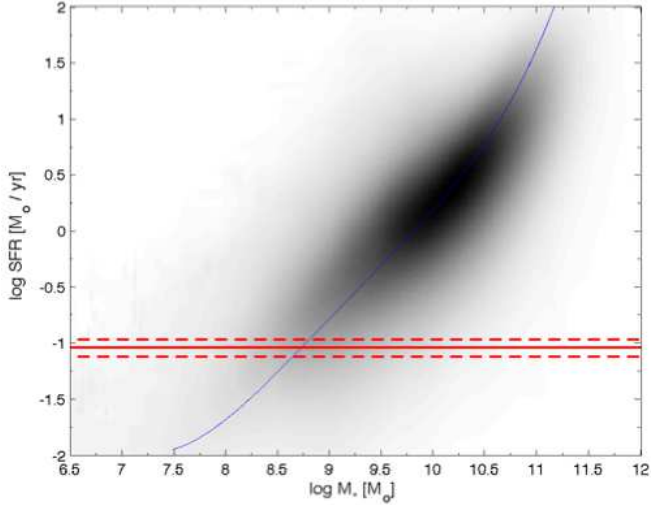}
\end{center}
\caption{SFR versus total stellar mass of star-forming galaxies from the SDSS (Brinchmann et al. 2004). A polynomial-fit fiducial to those data is shown in blue. 
Our measurement for J021904 and the 1$\sigma$-uncertainties are 
indicated by red lines.}
\end{figure}

However, this result should be taken with caution, since 
masses obtained this way should in principle 
be corrected for aperture effects that account for  differences between the total galaxy luminosity and the flux caught in the fibre (SDSS) or slit (MODS), respectively. 
For instance, at the median redshift  of 0.1 of  the SDSS galaxies (Blanton et al. 2003; Brinchmann et al. 2004), the SDSS fibres only sample $\sim$ 1/3 of the galaxy light. 
A possible bias can then be introduced if radial gradients in the galaxy properties are present, predominantly in the M/L ratio. 
While Brinchmann et al. (2004) carefully corrected their large reference sample, we did not further investigate any such aperture effects and will rather 
resort to more reliable, independent mass indicators in the following. 

Finally, we note that, if star formation had been enhanced during the merger event, the actual mass of the unperturbed galaxy would be overestimated 
with respect to its intrinsically lower SFR value. 

\vspace{1ex}
{\em (2) Stellar mass from metallicity:} 
As above, we interpolated the mass-metallicity relation defined by the SDSS sample of Brinchmann et al. (2004) and Tremonti et al. (2004; their Eq.~3)  to 
the oxygen abundance of J021904.
Since Tremonti et al. used the R$_{23}$ indicator as a measure of the oxygen abundance of their sample, for consistency, 
we used here the abundance of J021904 from the same index (see red lines in Fig.~9). 
\begin{figure}[tb]
\begin{center}
\includegraphics[angle=0,width=1\hsize]{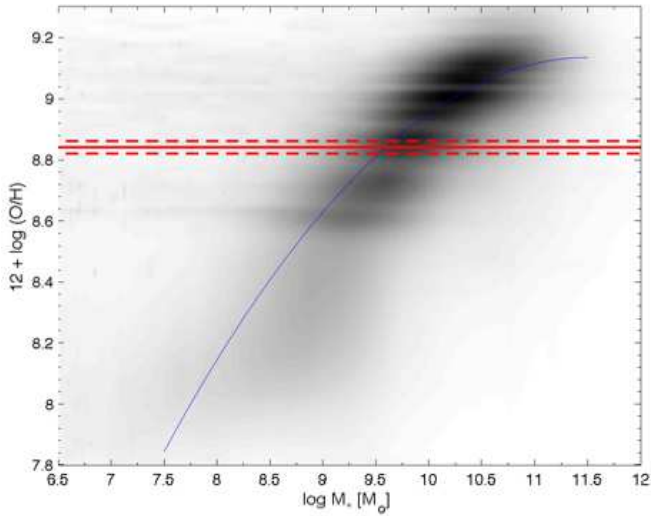}
\end{center}
\caption{Same as Fig.~8, but for the oxygen abundances from Tremonti et al. (2004). The blue line is the second-order polynomial fit to the relation
suggested by those authors.}
\end{figure}
This yields a stellar mass for the galaxy of M$_{\ast}$=(3.9$^{+0.6}_{-0.5}$)$\times$10$^9$ M$_{\odot}$, which is considerably  larger than 
 the above estimate from the SFR. 
However, this estimate is deemed more reliable since the oxygen abundance does not suffer from 
the biasses due to aperture effects discussed above. 

\vspace{1ex}
{\em (3) Fundamental plane:} 
By combining all of the above parameters of SFR, metallicity (taken as the oxygen abundance), and stellar mass, Lara-L\'opez et al. (2010) introduced a 
fundamental plane for ``main-sequence'' star-forming  galaxies in the field that was based on the SDSS sample we also employed for comparisons above. 
For J021904, this yields  M$_{\ast}$=(2.1$^{+1.2}_{-0.8}$)$\times$10$^9$ M$_{\odot}$. 
As for the estimate from the oxygen abundance alone, this result is more 
reliable compared to the low, SFR-based value, since there are no dependencies  on aperture effects.
Furthermore,  as Lara-L\'opez et al. (2010) show, there is  no evolution of  the fundamental plane with redshift, while there may be such trends within the 
 individual relations (Figs.~7,8; Erb et al. 2006). 

\vspace{1ex}
{\em (4) Spectral energy distribution (SED) fitting:}
Given the galaxy's faintness (at $g$=19.6 mag) no photometric information beyond the SDSS magnitudes is available in any other survey such as 
in the UV (GALEX) or the infrared (e.g., 2MASS or IRAS). 
Here we used the  MAGPHYS  code (da Cunha et al. 2008), which employs the stellar population synthesis models of Bruzual \& Charlot (2003) 
to interpret galaxy SEDs in terms of stellar parameters, also accounting for the properties of the galaxies' dust and interstellar extinction. 
While emission line spectra are not considered, we obtain a best-fit of the observed SED (Fig.~10) of J021904 for a template with a 
stellar mass of  M$_{\ast}$=(4.3$^{+1.5}_{-1.0}$)$\times$10$^9$ M$_{\odot}$, in line with the estimates from its metallicity and the fundamental relations
discussed above. Furthermore, the inferred template  metallicity of $Z/Z_{\odot}$ of 0.62 is  in line with our metal-rich value from the O/H measurements. 
\begin{figure}[htb]
\begin{center}
\includegraphics[angle=0,width=1\hsize]{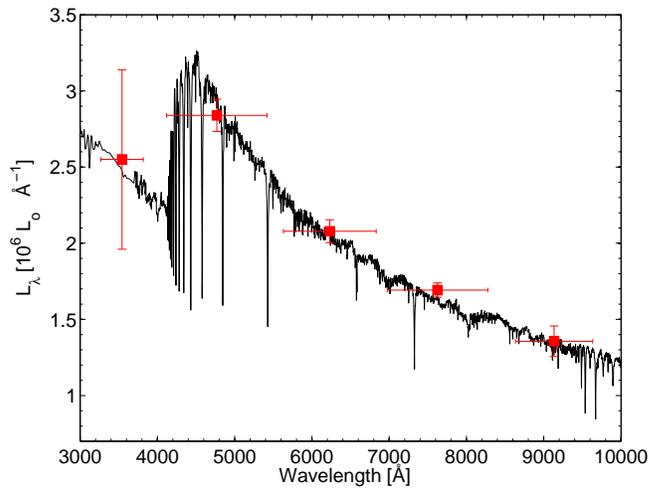}
\end{center}
\caption{Best-fit SED from the Bruzual \& Charlot (2003) models, obtained with  MAGPHYS (black). The observed SDSS fluxes are shown in red, where 
horizontal error bars reflect the filter widths.}
\end{figure}

In conclusion, we find a mean stellar mass of J021904 from all four estimates of
M$_{\ast}$=(3.1$^{+1.2}_{-2.5}$)$\times$10$^9$ M$_{\odot}$, consistent with the 
typical mass of the LMC and star-forming dwarf galaxies in the Local Universe (van der Marel et al. 2002). 
\section{Overall characterization of J021904 -- an LMC analog?}
Depending on the filter used, we found an { exponential effective, or half-light, radius of J021904 on the order of 8 kpc}, in excellent agreement with the Petrosian radius listed in the SDSS.
Coupled with its stellar mass and/or absolute magnitude estimate, this places this galaxy  in the realm of (dwarf) elliptical and Virgo cluster galaxies
in the effective radius-mass/magnitude diagrams of, e.g., Misgeld and Hilker (2011; their Figs.~1,4). A deviation of J021904 in these scaling relations towards larger radii can be ascribed to its irregular, merger-induced 
morphology and the possible tidal features. 
This compares to a smaller  (exponential) radius of 0.7--1.4 kpc of the LMC, which differs with the age of the stellar populations, 
and its tidal radius of $\sim$15 kpc (van der Marel 2006; Balbinot et al. 2015). In turn, the LMC's  H\,{\sc i} disk has an extent od 6.5 kpc.

J021904 is characterized by a SFR of 0.09 M$_{\odot}$\,yr$^{-1}$. 
This compares with the 
average,  {\em global} SFR of the  LMC of 0.2 M$_{\odot}$\,yr$^{-1}$, which varied, however, within a  factor of two over its history (Harris \& Zaritsky 2009). 
Thus the LMC saw extended, quiescent phases 5--12 Gyr ago, where stars formed at a comparable order of magnitude as seen in J021904. 
Additionally, it experienced repeated periods of enhanced star formation within the last 3 Gyr, during which the metallicity of the LMC increased by a factor of two (see also 
Dopita et al. 1997). These peaks  were 
likely triggered by close interactions with the SMC. 
While the moderate SFR found for J021904 is not unambiguously indicative of a merger-induced star formation activity, 
the double-cored morphology and emergence of apparent tidal features could suggest that 
J021904 is currently undergoing a merger within a LMC-SMC-type system.

The metallicity of the LMC has been determined from a large amount of various tracers. 
The stellar metallicity distribution indicates a metal-poor mean [Fe/H] of $-0.6$ dex (Cole et al. 2000) with a tail down to 
approximately $-1.8$ dex (e.g., Hill et al. 2000). Given the typical [O/Fe] enhancements of the more metal-rich 
stars in dwarf spheroidal galaxies in the MW to Solar values and higher enhancements to $\sim$0.4 dex at the metal-poor end 
(e.g., Gilmore \& Wyse 1991; Shetrone et al. 2001; Koch 2009; Hendricks et al. 2014) the entire covered range would translate to an oxygen abundance on 
the order of 7.8 to 8.3 dex.
Similarly, oxygen abundances of  planetary nebulae have been determined as 8.10$\pm$0.25 dex (Stasi\'nska et al. 1998), in line with 
measurements in H\,{\sc II} regions of 8.35$\pm$0.06 dex (Russell \& Dopita 1992).
Moreover, the prominent Magellanic Stream has been characterized as a  metal-poor component, at 12+log(O/H) = 7.46--7.66  dex (Fox et al. 2013). 
In comparison, J021904 is a metal-rich system that lies above the nebular abundances for the LMC by at least 0.3 dex, which could indicate that 
the potential merger happened with an already metal-rich system.

Using three independent, reliable methods, viz. the mass-metallicity relation, the fundamental plane for main-sequence galaxies, and 
from SED-fitting, we estimated the stellar mass of J021904 to be 3.1$\times$10$^{9}$ M$_{\odot}$. 
Considering the large uncertainties on our measurement, this value is in excellent agreement with the total stellar mass of the LMC of 
M$_{\ast, {\rm LMC}}$=2.9$\times$10$^{9}$ M$_{\odot}$ (van der Marel et al. 2002). 

The line asymmetries seen in the spectrum of J021904 could be due to several dynamic processes, 
while it is unlikely that we are sampling the depth extent of this object: at the redshift of  the galaxy, the split of 140 km\,s$^{-1}$ corresponds to $\sim$2.1 Mpc
which is much larger than the apparent, radial  extent of the system on the order of a few tens of kpc. 
One viable cause for the splits could be rotation. 
Both our measured v$_{\rm rot}$ and the inferred mass are in fact consistent with the stellar Tully-Fisher relation observed in local galaxy samples 
(e.g., Kassin et al. 2007; Miller et al. 2011) and also predicted by simulations of galaxy mergers  (Covington et al. 2010).
One drawback of this scenario is that due to the orientation of the MODS slit (Fig.~3), the observed line split would imply  rotation
around the major axis or a significant inclination of the system, all of which is feasible if one was to consider the exact geometry of the infalling satellite's orbit.
Clearly, future long-slit or IFU observations are needed to settle the relative dynamics of J021904.
An alternative could be galactic outflows or spatially separated star forming regions, as suggested by Pilyugin et al. (2012).
Finally, this split could also simply reflect the relative velocities of the merging components
on an order of magnitude similar to the (past) relative velocities of the Magellanic clouds. 
\section{Discussions and conclusions}
Our discovery of J021904 as an irregular, moderately star-forming, Magellanic galaxy with a complex, { visual} structure 
{ brings forth another contender for 
a major merger event
that already acts in the regime of dwarf galaxy masses}
(see also Rich et al. 2012). In terms of their star-forming properties, Overzier et al. (2008) note that mergers of gas-rich, low-mass ($\sim 10^{10}$ M$_{\odot}$) systems 
are important triggers of intense star formation activity. 

Besla et al. (2012) suggested that many of the LMC structures such as its warped stellar bar are a natural outcome of a head-on collision with the lower-mass SMC.
These features are an important trademark of the  Magellanic Irregulars, indicating that their morphological evolution has indeed been significantly shaped by 
similar major mergers on the dwarf galaxy scale. 

Based on a large sample of galaxy pairs,  Davies et al. (2015)  found  that star formation is enhanced in major mergers and in the primaries of minor mergers, while it is 
suppressed in the satellites of minor mergers. 
In fact, our observed line-splitting and its morphology indicate that J021904 contains two star-forming components, revealing a major merger with the bluer, Western component we analysed 
in this work being the primary, accreting partner.  
When inferring masses from SFRs, one should then 
bear in mind that ÓenhancedÓ  star-formation does strictly not only refer to ``star-forming'' per se, but that the galaxy is forming stars more actively than non-merging galaxies of 
comparable mass.

One open question thus remains if J021904 has evolved as an isolated system or whether it is part of a larger 
structure such as a galaxy group or cluster. For instance, Tollerud et al. (2011) estimated that  12\% of MW-type galaxies host LMC-type satellites within 75 kpc, which corresponds to 34$\arcsec$ at the distance of 
our galaxy. 
For comparison, we thus turn to the Local Group with its (zero-velocity) radius of $\sim$1 Mpc (Karachentsev et al. 2009), which 
corresponds to 7.6$\arcmin$ at the distance of J021904.
This is essentially comparable to the entire field of view of the MIMO camera (Sect.~2).  
Consulting the SDSS, we identified eight candidates with photometric redshifts that would place them within a projected radius of 2 Mpc of our Magellanic candidate within their respective  errors (note that those are typically on the order of 50\%). 
However, none of them fall within the footprint of our MIMO imaging and the SDSS images of those faint objects indicate no signs of interaction, nor are any of them extended L$^*$ galaxies 
so that we conclude that J021904 is most  likely an isolated, star-forming field galaxy. 
\acknowledgments
AK  and MJF gratefully acknowledge the Deutsche Forschungsgemeinschaft for funding from  Emmy-Noether grant  Ko 4161/1. 
We thank V. Belokurov and M. Walker for kindly providing their MMT images of this region and D.J. Watson, T. Lisker, G. Lamer, and J. Hennawi for helpful discussions. 
The anonymous referee is thanked for a very swift and helpful report, prompting the CAS analysis. 
Based in parts on observations at Kitt Peak National Observatory, National Optical Astronomy Observatory (NOAO Prop. ID:2010B-0588; PI: R.M. Rich), 
which is operated by the Association of Universities for Research in Astronomy (AURA) under a cooperative agreement with the National Science Foundation. 
This paper also used data obtained with the MODS spectrographs built with
funding from NSF grant AST-9987045 and the NSF Telescope System
Instrumentation Program (TSIP), with additional funds from the Ohio
Board of Regents and the Ohio State University Office of Research.
This paper made use of the modsIDL spectral data reduction reduction pipeline developed in part with funds provided by NSF Grant AST-1108693.
\end{document}